# Concentration-modulation FT emission spectroscopy of TiCl$_4$/He plasma. Analysis of the $C\ ^4\Delta$- $X\ ^4\Phi$ $\Delta v=0$ perturbed transitions of TiCl.

**Hervé Herbin, Robert Farrenq, Guy Guelachvili, Nathalie Picqué**
*Laboratoire de Photophysique Moléculaire, Unité Propre du C.N.R.S., Bâtiment 350, Université de Paris-Sud, 91405 Orsay, France*
guy.guelachvili@u-psud.fr     nathalie.picque@u-psud.fr     http://www.laser-fts.org

**Abstract:** A TiCl$_4$/He plasma is observed by high resolution double-modulation FTS using concentration-modulation as a selective detection method. Analysis of the $C^4\Delta$-$X^4\Phi$ $\Delta v=0$ transitions of $^{48}$Ti$^{35}$Cl reveals perturbations affecting the $C\ ^4\Delta_{1/2}$ sub-state.
©2005 Optical Society of America

One major step in the development of infrared spectroscopy of unstable species has been the introduction of selective detection techniques like concentration modulation [1] or velocity modulation [2] laser spectroscopies of discharge plasmas. These methods exploit physical properties of unstable molecules like lifetime, electrical charge, paramagnetism to distinguish them against the more abundant precursors. Selective detection techniques in FTS have already been demonstrated with step-scan interferometers [3-5] and with rapid-scan interferometers [6,7]. They have never been extensively taken advantage of, most likely because of their instrumental complexity. In this paper, high resolution double modulation Fourier-transform spectroscopy has been applied to selective detection of short-lived species by concentration modulation in a TiCl$_4$/He plasma. Emission spectra of the $C^4\Delta$-$X^4\Phi$ $\Delta v=0$ rovibronic transitions have been recorded and analyzed.

The principle of double-modulation FTS has been described elsewhere [4,8]. Here it is demonstrated using concentration modulation as a selective modulation. The modulation originates from an alternating current discharge, with synchronous detection of the interferogram at twice the discharge frequency. Two selective interferograms are detected in quadrature on two separate channels. The spectra are obtained from the emission of a glow discharge through a continuous flow of TiCl$_4$ and He. The discharge tube is 25 cm long and has an inner diameter of 0.7 cm. Helium (1300 Pa) and traces of TiCl$_4$ vapor enter the cell by a central inlet. A (800 V, 0.3 A, 10 kHz) *ac* power supply, driven by an audio amplifier connected to a transformer, is connected to the two electrodes of the tube. The emission of the source is sent into a stepping-mode Fourier transform interferometer. The interferometer is equipped with a CaF$_2$ beamsplitter and two liquid-nitrogen cooled InSb detectors. The recording spectral range is about 1800-4000 cm$^{-1}$ and the non-apodized resolution equal to about $10\ 10^{-3}$ cm$^{-1}$.

Fig. 1 and Fig. 2 display small portions of spectrum (nb.3525). Fig. 1 illustrates that the technique provides effective selectivity against stable precursor emission, since the lines of HCl, which is present as an impurity, are removed from the selective spectra. Fig.2 shows that the thermal background and the superimposed absorption lines due to atmospheric CO$_2$ present in the nonselective spectrum are also eliminated from the selective spectra.

The spectra have also allowed the observation of the $C\ ^4\Delta$- $X\ ^4\Phi$ $\Delta v=0$ transitions of $^{48}$Ti$^{35}$Cl, with $v=$ 0,1,2,3, around 3 µm. The 0-0 rovibronic transitions had already been investigated by Ram and Bernath [9]. In the work reported here, the rovibronic transitions 1-1, 2-2, 3-3 and weak perturbations occurring in the 0-0, 1-1, and 2-2 $C\ ^4\Delta_{1/2}$ - $X\ ^4\Phi_{3/2}$ subbands, that had not been reported previously, have been identified

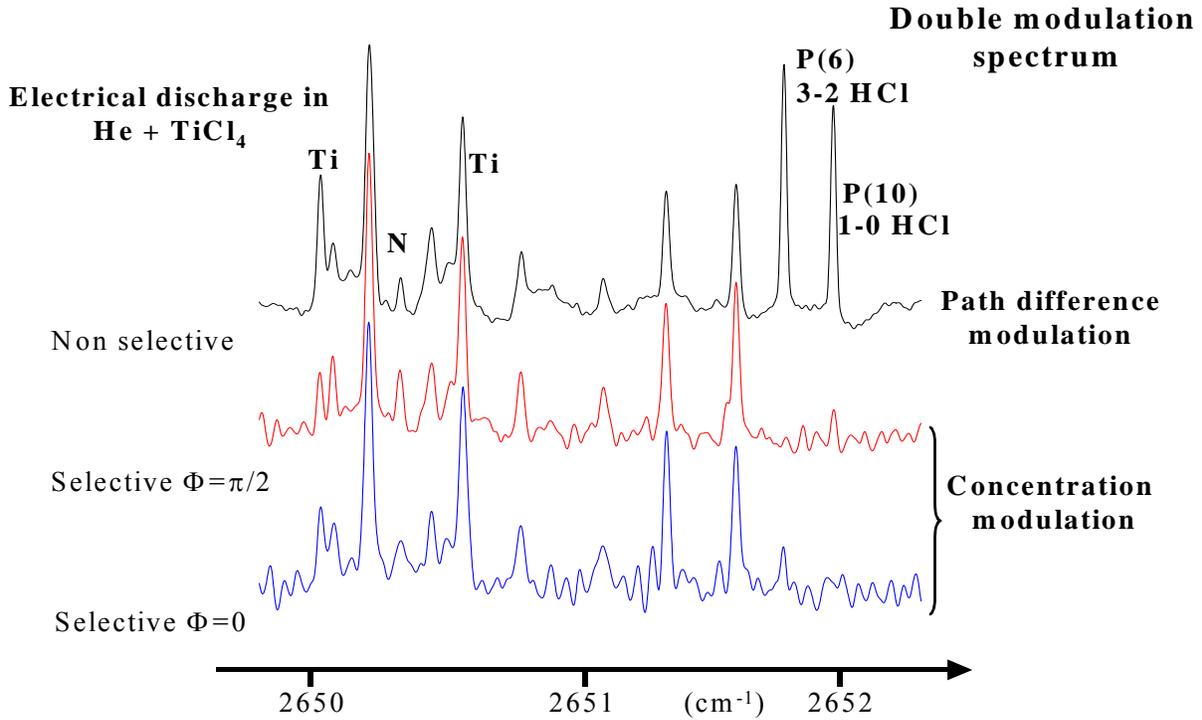

Fig. 1: Nonselective and selective portions of a high-resolution spectrum: lines from HCl stable molecule are removed from the selective components. The two selective spectra result from the synchronous detection of the selective interferogram, which is made in phase and in quadrature on two separate channels

**Double modulation spectrum with concentration modulation**

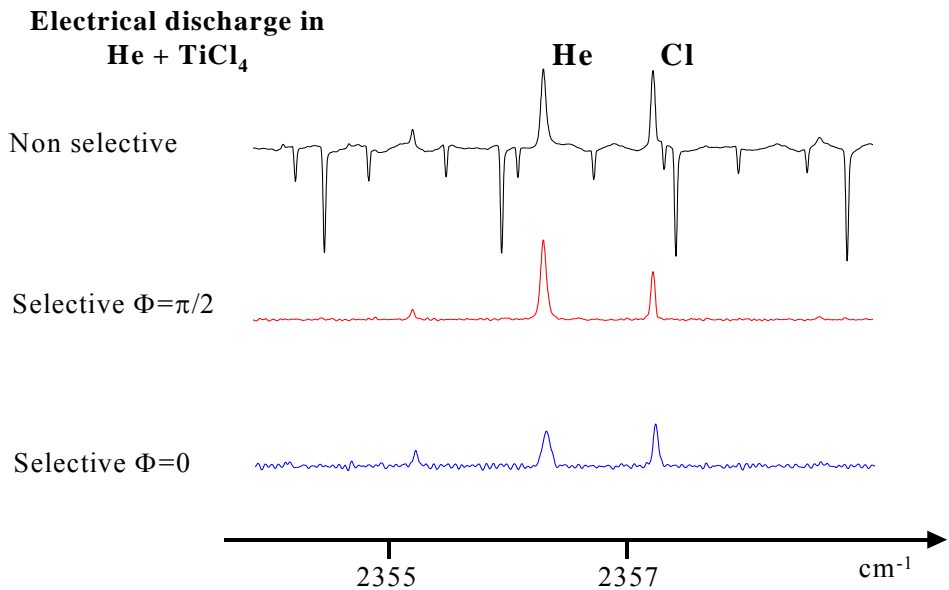

Fig. 2: Illustration of the selectivity : thermal background and atmospheric absorption lines from the $\nu_3$ and $\nu_2+\nu_3-\nu_2$ bands of $CO_2$ appear only on the nonselective component.

and analyzed. The perturbations are due to $\Lambda$-doubling and avoided crossing phenomena induced by the never observed $B\ ^4\Pi$ and $A\ ^4\Sigma^-$ states. Effective analysis of the unperturbed and perturbed bands has been performed, as well as a global analysis, including the accurate submillimeter lines measured by Ref.[10]. The equilibrium structure of $^{48}$Ti$^{35}$Cl in the $C\ ^4\Delta$ state is determined from the resulting molecular constants [11].

This research was financially supported by the Programme National de Physique et Chimie du Milieu Interstellaire (PCMI) du Centre National de la Recherche Scientifique (CNRS).

**References**
[1] T. Amano, "The $\nu_1$ fundamental band of HCO$^+$ by difference frequency laser spectroscopy", *Journal of Chemical Physics* **79**, 3595-3595 (1983).
[2] C.S. Gudeman, M.H. Begeman, J.Pfaff, R.J. Saykally, "Velocity modulated infrared laser spectroscopy of molecular ions: the $\nu_1$ band of HCO$^+$", *Physical Review Letters* **50**, 727-731 (1983).
[3] G. Guelachvili, "Selective detection of paramagnetic species by high-information Fourier-transform spectrometry", *J. Opt. Soc. Amer. B* **3**, 1718-1721 (1986).
[4] N. Picqué, G. Guelachvili, "High resolution multi-modulation Fourier transform spectroscopy", *Applied Optics* **38**, 1224-1230 (1999) and references therein
[5] N. Picqué, "Fast phase-selective detection of transient species with step-scan Fourier transform spectroscopy", *Journal of the Optical Society of America B* **19**, 1706-1710 (2002) and references therein.
[6] T. Imajo, S. Inui, K. Tanaka, T. Tanaka, "Interferogram amplitude modulation technique for selective detection of transient species with a continuous-scan Fourier-transform spectrometer", *Chemical Physics Letters* **274**, 99-105, 1997.
[7] X. Hong and T.A. Miller, "Velocity modulated Fourier transform emission as a plasma diagnostic and a spectroscopic tool", *Journal of Chemical Physics* **101**, 4572-4577, 1994.
[8] H. Herbin, R. Farrenq, G. Guelachvili, N. Picqué, "Doppler-shifted transitions of a neutral molecule revealed by velocity modulation FTS", contributed paper in the present conference.
[9] R. S. Ram, P. F. Bernath, "Fourier transform infrared emission spectroscopy of the $C^4\Delta$-$X^4\Phi$, $G^4\Phi$-$X^4\Phi$, and $G^4\Phi$-$C^4\Delta$ systems of TiCl", J. Mol. Spectrosc. **186**, 113-130 (1997).
[10] A. Maeda, T. Hirao, P. F. Bernath, T. Amano, "Submillimeter-Wave Spectroscopy of TiCl in the ground electronic state", *J. Mol. Spectrosc.* **210**, 250-257 (2001).
[11] H. Herbin, R. Farrenq, G. Guelachvili, B. Pinchemel, N. Picqué, "Perturbation analysis in the $X\ ^4\Phi$ - $C\ ^4\Delta$ rovibronic transitions of $^{48}$Ti$^{35}$Cl at 3 µm", *J. Mol. Spectrosc.* **226**, 103–111 (2004).